\documentclass[twocolumn]{aastex63}
\usepackage[caption=false]{subfig}

\usepackage{savesym}
\savesymbol{tablenum}
\usepackage{siunitx}
\restoresymbol{SIX}{tablenum}
\submitjournal{ApJ Letters}
\accepted{July 28, 2020}
\shortauthors{N. Trueba et al.}
\shorttitle{The Redshifted Inner Disk Atmosphere in 4U 1916$-$053}

\begin{document}

\title{A Redshifted Inner Disk Atmosphere and Transient Absorbers in the Ultra-Compact Neutron Star X-ray Binary 4U 1916$-$053}

\correspondingauthor{Nicolas Trueba}
\email{ntrueba@umich.edu}

\author{Nicolas Trueba}
\affiliation{Department of Astronomy, University of Michigan, Ann Arbor, MI}
\author{J.~M.~Miller}
\affiliation{Department of Astronomy, University of Michigan, Ann Arbor, MI}
\author{A.~C.~Fabian}
\affiliation{Institute of Astronomy, University of Cambridge, Madingley Road, Cambridge CB3 OHA, UK}
\author{J.~Kaastra}
\affiliation{SRON, Netherlands Institute for Space Research, Sorbonnelaan 2, 3584 CA Utrecht, The Netherlands}
\author{T.~Kallman}
\affiliation{NASA Goddard Space Flight Center, Code 662, Greedbelt, MD 20771, USA}
\author{A.~Lohfink}
\affiliation{Department of Physics, Montana State University, Bozeman, MT 59717-3840, USA}
\author{D.~Proga}
\affiliation{Department of Physics, University of Nevada, Las Vegas, Las Vegas, NV 89154, USA}
\author{J.~Raymond}
\affiliation{Harvard-Smithsonian Center for Astrophysics, 60 Garden Street, Cambridge, MA 02138, USA}
\author{C.~Reynolds}
\affiliation{Institute of Astronomy, University of Cambridge, Madingley Road, Cambridge CB3 OHA, UK}
\author{M.~Reynolds}
\affiliation{Department of Astronomy, University of Michigan, Ann Arbor, MI}
\author{A.~Zoghbi}
\affiliation{Department of Astronomy, University of Michigan, Ann Arbor, MI}

\begin{abstract}
The very small accretion disks in ultra-compact X-ray binaries (UCXBs) are special laboratories in which to study disk accretion and outflows. 
We report on three sets of new (250 ks total) and archival (50 ks) Chandra/HETG observations of the ``dipping'' neutron-star X-ray binary 4U 1916$-$053, which has an orbital period of $P\simeq 50$~minutes.  We find that the bulk of the absorption in all three spectra originates in a disk atmosphere that is redshifted by $v\simeq 220-290$ $\text{km}$ $\text{s}^{-1}$, corresponding to the gravitational redshift at radius of $R \sim 1200$ $GM/{c}^{2}$.  This shift is present in the strongest, most highly ionized lines (Si XIV and Fe XXVI), with a significance of 5$\sigma$. Absorption lines observed during dipping events (typically associated with the outermost disk) instead display no velocity shifts and serve as a local standard of rest, suggesting that the redshift is intrinsic to an inner disk atmosphere and not due to radial motion in the galaxy or a kick. In two spectra, there is also evidence of a more strongly redshifted component that would correspond to a disk atmosphere at $R \sim 70$ $GM/{c}^{2}$; this component is significant at the 3$\sigma$ level.  Finally, in one spectrum, we find evidence of disk wind with a blue shift of $v = {-1700}^{+1700}_{-1200}$ $\text{km}$ $\text{s}^{-1}$. If real, this wind would require magnetic driving.

\end{abstract}

\section{Introduction}\label{sec:intro}

Compact, short-period X-ray binaries are remarkable systems, but 4U 1916$-$053 is special even within this class. Whereas many short-period binaries are faint, 4U 1916$-$053 is relatively bright, typically exhibiting an X-ray flux of $F\geq 0.2$~Crab in soft X-rays \citep[e.g.,][]{Galloway2008}.  This fact, and the low interstellar column density along its line of sight  ($N_{H}\simeq 2.3\times 10^{21}~{\rm cm}^{-2}$; \citealp{ISM2016}), facilitate detailed spectroscopy and timing studies of 4U 1916$-$053 \citep[e.g.,][]{DiazTrigo2006,Iaria2015,Gambino2019}.

%-----------------------------------------------------------------------------
\begin{figure*}
\centering
\includegraphics[width=0.99\textwidth,angle=0]{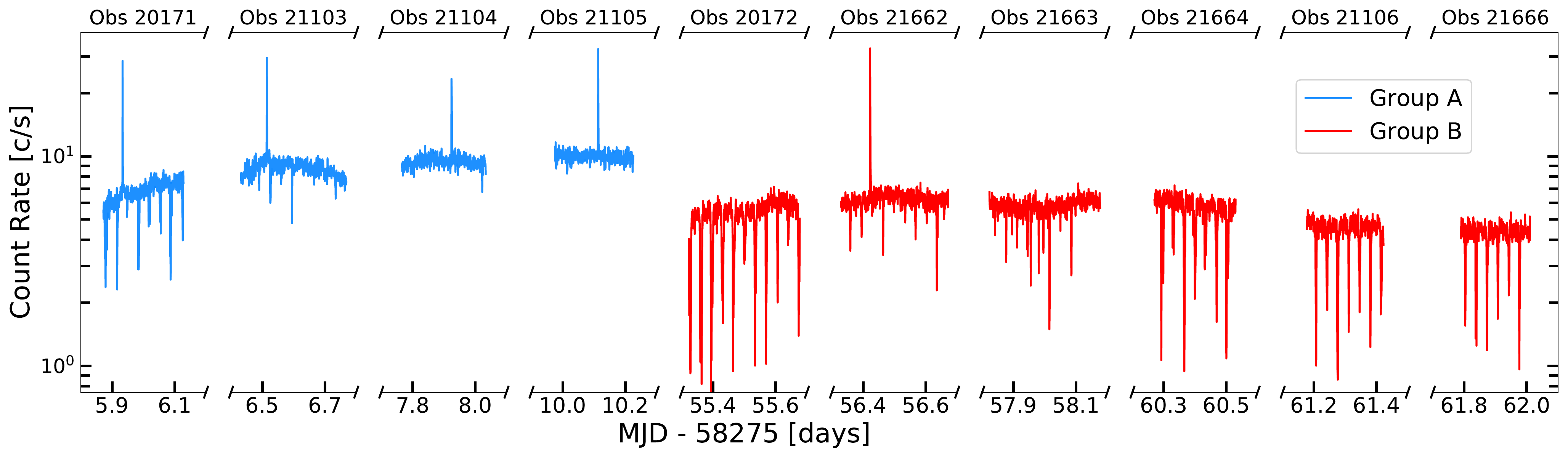}
\figcaption[t]{\footnotesize \label{fig:lc} Barycenter-corrected Chandra/HETG lightcurve for the 10 observations of 4U 1916$-$053. The spectra where grouped (Group A and Group B), based on their relative intensity and the $~$45-day gap between the two groups.}
\end{figure*}
%-----------------------------------------------------------------------------

The detection of Type-1 X-ray bursts \citep[e.g.,][]{Galloway2008}, and coherent oscillations within the bursts (Galloway et al.\ 2001), clearly signal that the binary harbors a neutron star.  4U 1916$-$053 does not exhibit eclipses, but the presence of strong dips in its X-ray light curve \citep{Walter1982}  -- likely associated with structures in the outer disk \citep[e.g.,][]{DiazTrigo2006} -- signals that it is viewed at a high inclination.  These dips, and optical properties, point to an orbital period of just $P \simeq 50$~minutes \citep{White1982,Walter1982}.  Given these system parameters, it is likely that the companion is degenerate ($M = 0.065\pm 0.01~M_{\odot}$, assuming a $M = 1.4~M_{\odot}$ primary; \citealp{Heinke2013}), and that the accretion flow is dominated by helium rather than hydrogen \citep[e.g.,][]{Joss1978}.  This composition is confirmed by optical spectroscopy \citep[e.g.,][]{Nelemans2006}.

4U 1916$-$053 has been extensively studied with modern X-ray telescopes.  The detection of ionized absorption lines, including He-like Fe XXV and H-like Fe XXVI, was first reported in an analysis of {\it XMM-Newton} data \citep{Boirin2004}.  
A later study with the {\it Chandra}/HETGS found no credible velocity shift, even in these most ionized lines, and associated the lines with the extreme outer disk \citep{Juett2006}.  A more comprehensive study of ionized absorption in X-ray ``dippers'' concluded that variations in ionized absorbing regions in the outer disk cause the dipping phenomenon \citep{DiazTrigo2006}.  A recent analysis of archival {\it Suzaku} data also concluded that the absorption is associated with the outer accretion disk \citep{Gambino2019}.

It is not clear if ionized but static disk atmospheres are related to blue-shifted disk winds, or if they are disconnected.  It is possible that such atmospheres turn into winds when certain physical conditions are met.  In GRS 1915$+$105, for instance, ionized absorption is sometimes evident in the Fe K band with no significant blue-shift \citep[e.g.,][]{Lee2002, Neilsen2019}.  In other phases, a disk wind is clearly detected \citep[e.g.,][]{Ueda2009, Neilsen2009}; indeed, the wind can be as fast as $v/c = 0.03$ \citep{Miller2016,Zoghbi2016}.  It is particularly interesting to search for winds and atmosphere--wind connections in compact binaries, since even the outer disk may be too small to launch thermal disk winds.  

The goals of better understanding disk atmospheres, winds, and wind driving mechanisms motivated us to observe 4U 1916$-$053 for an additional 250~ks in {\it Chandra} Cycle 19.  In Section 2, we describe the observations and data reduction process.  Section 3 details the analysis procedure that we employed, and the outcomes.  In Section 4, we discuss our results in the context of other recent studies of disk atmospheres and winds.

\section{Observations and Data Reduction}\label{sec:data}

4U 1916$-$053 has been observed with the Chandra/HETG on eleven occasions: once in 2005 (ObsID 4584, 46 ks) plus ten observations in 2018 (totalling 250 ks). The source was faint enough during all observations for the exposures to be conducted in the ACIS TE (Timed Exposure) mode and FAINT data mode. 

The data from all eleven observations were reduced using CIAO version 4.7 and CALDB version 4.7.6. A barycenter correction was applied to the data using the {\it axbary} routine. The CIAO routines {\it tg\_findzo}, {\it tg\_create\_mask}\footnote[1]{The width\_factor\_hetg parameter in this routine was set to 10 in order to reduce contamination from MEG photons onto the HEG, thereby greatly increasing the sensitivity of the HEG above the FeK band.}, {\it tg\_resolve\_events}, {\it tgextract}, {\it mkgrmf}, and {\it fullgarf} were used to extract the first-order HEG and MEG spectrum, redistribution matrix, and ancillary response files for each observation. 
Plus and minus orders were combined using {\it combine\_grating\_spectra}. Time filtering of spectra (including the removal of dips and bursts events, as well obtaining phase-resolved spectra) was done using the {\it dmcopy} routine. 

Lightcurves for the ten 2018 observations of 4U 1916$-$053 are shown Figure \ref{fig:lc}. 
Each successive observation is separated by $1-3$ days except for a $\sim 45$ day gap, separating the data set in two epochs. 
Visually, the lightcurves suggest that the two sets represent two distinct luminosity states.
This is also apparent when examining the 2$-$9 keV continua for all observations (including the archival HETG observation, ObsID 4584): The continuum becomes harder as the luminosity increases, ranging from the least-luminous and softest Group B observations to the brightest and hardest spectrum of ObsID 4584.

An examination of the extracted spectra reveals the presence of ionized, H-like absorption lines in the Fe K band and at $\sim 2$ keV (Si XIV). The low signal-to-noise ratio in individual spectra, however,  makes it impossible to extract useful information for each observation. The continuum shapes for observations within each of the two sets show only minor discrepancies, therefore we combined the spectra of each set to produce two combined spectra corresponding to each epoch. We will refer to these combined spectra as Spectrum A and Spectrum B. 
Although the continuum of ObsID 4584 is not too dissimilar to the continua of the observations in Spectrum A (comparable spectral shape, but the former is noticeably more luminous), the observations are separated by $\sim$14 years and therefore the absorption in the system might be different.  We therefore chose to analyze this observation separately and refer to it as Spectrum $\Gamma$. The raw, low S/N spectra of individual observations were combined using the standard Gehrels errors delivered by the standard CIAO pipeline. Once combined, we replaced the errors in the average spectra with Poissonian errors in order to more accurately represent the errors once the spectra were binned.

The light curves also display both dipping and burst events. Although these contribute very little information to the integrated spectrum for each observation, these events were initially removed entirely when extracting each spectrum in an effort to remove any line absorption originating from the dips. The dips display complexity in their geometry: the shape, depth, and duration of dips changes between and within each observation, as well as the occasional appearance of secondary dips at roughly twice the orbital frequency. Any detailed analysis on the nature of the structure producing the dips is beyond the scope of this work. However, regardless of shape, all dips have clearly defined starting and ending points, and the lack of any orbital modulation other than discrete dips strongly suggests that any absorption coming from the outer disk is confined to the dipping events.

\section{Analysis \& Results}\label{sec:analysis}

Spectral modeling was performed using SPEX version 3.05.00 \citep{Kaastra1996,Kaastra2018} and SPEXACT version 3.05.00 atomic database and associated routines. 
The data were binned when necessary using the ``obin'' command, which rebins a spectrum to the optimal bin size given the particular statistics of a given source.
Continuum fits were performed in the 1.9 to 9.0 keV range. Each individual HETG spectrum was a composite of their MEG and HEG spectra, where 
the higher resolution HEG made up the 3.8 to 9.0 keV range, while the MEG covered the 1.9 to 3.0 keV portion due to its higher effective area at low energies. 
There are no lines of significance in the 3.0 to 3.8 portion of any of the 3 spectra.  

The continuum for all three data sets was modeled with a blackbody (bb in SPEX) plus a disk blackbody (dbb), modified by interstellar absorption (absm).
The neutral absorbing column is low and was fixed at ${N}_{H} = 4 \times 10^{21}~{\rm cm}^{-2}$, broadly consistent with the aforementioned $2.3\times 10^{21}~{\rm cm}^{-2}$ value \citep{ISM2016} after accounting for differences in abundances and models. 
The best-fit temperatures and normalization parameters for the blackbody and disk blackbody components are within the acceptable ranges for these parameters,
suggesting a neutron star temperatures of 2.2, 1.8, and 1.7 keV, and a disk temperatures of 0.9, 0.8, and 0.9 keV for Spectrum $\Gamma$, A, and B, respectively.

The residuals of the best-fit continua reveal similar ionized absorption features among the 3 different spectra. Strong Fe XXVI and Si XIV absorption lines appear near their respective rest energies, likely originating in a disk atmosphere. The Fe XXVI line in Spectrum A, in particular, displays some complexity: a blue wing that could indicate the presence of a disk wind. The same complexity is perhaps also present in Spectrum B. In both spectra, however, this blue wing is only present in the profile of the Fe XXVI line. 
In addition, there is possible weak absorption feature in residuals for both Spectrum $\Gamma$ and Spectrum A at $\sim 6.8$ and $\sim 6.9$ keV, respectively. 
Although these features are relatively weak, the lack of instrumental features at these energies and their presence in two of our spectra suggests that 
these may be real absorption lines.  If real, it is unlikely these are blue-shifted Fe XXV absorption lines originating in a rapidly out-flowing gas given the lack of either an Fe XXVI or, alternatively, lower ionization lines (such as Ar XVI or Ca XX) at the same blueshift. Instead, these features could be highly redshifted Fe XXVI absorption, perhaps originating in a gravitationally redshifted absorber located near the compact object ($r < 100$ $GM/{c}^2$). In this case residuals at $\sim 6.6$ keV in Spectrum A could be the accompanying Fe XXV line complex. 

\begin{table*}[t]
\renewcommand{\arraystretch}{1.1}
\caption{Parameters for Best-Fit Absorption Models}
\vspace{-1.0\baselineskip}
\begin{footnotesize}

\begin{center}
\begin{tabular*}{\textwidth}{l c l @{\extracolsep{\fill}}  c  c   c c }
\tableline
\tableline

\multicolumn{7}{c}{Individual Spectra}\\
Spectrum & Model & Parameter & Zone 1 & Zone 2 & Zone 3 & $\chi^{2}/\nu$\\
\\ [-3.0ex]
\tableline 
\\ [-3.0ex]

Spectrum A&3 Zone&${N}_{He}$ (${10}^{22}$ ${\text{cm}}^{-2}$)&${50}^{\ddagger}_{-10}$ & ${10}_{-9\ddagger}^{+40\ddagger}$ & ${1.3}_{-1.1}^{+15.0}$   & 536/488 = 1.10\\
&& log $\xi$&${4.45}_{-0.05}^{+0.25}$ & ${4.80}_{-1.6}^{\ddagger}$ & ${3.8}_{-1.1}^{1.0\ddagger}$ & $-$ \\
&& ${v}_{abs}$ ($\text{km}$ $\text{s}^{-1}$) & ${230}_{-270}^{+500}$ & ${-1700}_{-1300}^{+1700\ddagger}$ & ${4200}_{-1600}^{+2900}$  & $-$\\
&& ${v}_{turb}$ ($\text{km}$ $\text{s}^{-1}$) & ${270}_{-220\ddagger}^{+230\ddagger}$ & ${300}^{\dagger}$ & ${1000}^{\dagger}$  &$-$ \\
&&${\sigma}_{emis}$ ($\text{km}$ $\text{s}^{-1}$)  &${5400}_{-4000}^{+15000}$ & $-$ & $-$ & $-$\\
&&${L}_{phot}$ (${10}^{36}$ erg/s) & 8.2 $\pm$ 2.2 & $-$ & $-$ & $-$ \\

\\ [-3.0ex]
\tableline 
\tableline 
\\ [-3.0ex]

Spectrum B &1 Zone &${N}_{He}$ (${10}^{22} {\text{cm}}^{-2}$)  &${33}_{-4}^{+47}$ & $-$ & $-$ & 491/493 = 0.996\\
& & log ${\xi}$ & ${4.4}_{-0.1}^{+0.4}$ & $-$ & $-$ & $-$ \\
&& ${v}_{abs}$ ($\text{km}$ $\text{s}^{-1}$) &${230}_{-130}^{+140}$ & $-$ & $-$ & $-$  \\
&& ${v}_{turb}$ ($\text{km}$ $\text{s}^{-1}$) &${150}_{-50}^{+80}$ & $-$ & $-$ & $-$  \\
&&${\sigma}_{emis}$ ($\text{km}$ $\text{s}^{-1}$) & $-$ & $-$ & $-$ & $-$ \\
&&${L}_{phot}$ (${10}^{36}$ erg/s) &  5.2 $\pm$ 1.5 & $-$ & $-$ & $-$\\

\\ [-3.0ex]
\tableline 
\tableline 
\\ [-3.0ex]

Spectrum $\Gamma$ &2 Zone &${N}_{He}$ (${10}^{22} {\text{cm}}^{-2}$)  &${34}_{-4}^{+25}$& $-$  & ${50}^{\dagger}$ & 482/486 = 0.99\\
&& log $\xi$&${4.25}_{-0.05}^{+0.20}$ &  $-$ &${5.3}_{-0.3}^{+0.7}$ & $-$  \\
&& ${v}_{abs}$ ($\text{km}$ $\text{s}^{-1}$) &${280} \pm 160$ & $-$ & ${6500}_{-1200}^{+1300}$  & $-$  \\
&& ${v}_{turb}$ ($\text{km}$ $\text{s}^{-1}$) & ${200}_{-70}^{+150}$ & $-$ &  ${1000}^{\dagger}$  &$-$ \\
&&${\sigma}_{emis}$ ($\text{km}$ $\text{s}^{-1}$) & ${9300} \pm 6000 $& $-$  & $-$ & $-$ \\
&&${L}_{phot}$ (${10}^{36}$ erg/s) & 10 $\pm$ 3 & $-$ & $-$ & $-$  \\

\\ [-3.0ex]
\tableline 
\tableline

\end{tabular*}
\vspace*{-1.0\baselineskip}~\\ \end{center} 
\tablecomments{Best-fit parameter values for the absorption model that best describes each spectrum, as described in the text. The parameters of absorber include the equivalent \emph{Helium} column (${N}_{He}$), the ionization parameter ($\xi$), the systematic velocity shift of the absorber (${v}_{abs}$), the turbulent velocity of the absorber (${v}_{turb}$), the dynamical broadening of the re-emission (${\sigma}_{emis}$), and the ionizing luminosity (from 13.6 eV to 13.6 keV). Quoted errors are at the $1\sigma$ level. The absorption in all spectra are dominated by a disk atmosphere (Zone 1): a \emph{nearly} static absorber which has fairly constant properties (${N}_{He}$, log~$\xi$, and ${v}_{abs}$) for all observations. Frozen parameters are marked with $\dagger$. Errors that reach the fitting range for the parameter before the required change in $\chi^{2}$ are marked with $\ddagger$. For Zone 2 and 3, ${v}_{turb}$ was kept frozen as these components are optically thin and not sensitive to line ratios. A large ${v}_{turb}$ of 1000 $\text{km}$ $\text{s}^{-1}$ was chosen that would approximately capture the width of the highly redshifted features (Zone 3) seen in Spectrum A and $\Gamma$. For Spectrum $\Gamma$, ${N}_{He}$ was frozen at $5\times$ ${10}^{23}$ ${\text{cm}}^{-2}$ (approximately the point at which a \emph{Helium} absorber becomes Compton-thick) as an alternative to the Compton-thin Zone 3 scenario in Spectrum A.}
\end{footnotesize}
\label{tab:fit}
\end{table*}

We modeled the line absorption using PION, a photoionized absorption model within SPEX. Unlike pre-calculated grid models, for instance those from XSTAR \citep{KallmanBautista} or CLOUDY \citep{Ferland2017}, PION self-consistently calculates the ionization balance of the absorber in real time, accounting for the changing shape of the ionizing continuum during the fitting process and as it is reprocessed by successive absorbers.  We note that PION models H-like lines as proper spin-orbit doublets, so any subtle shifts are not the result of unmodeled line structure.

%-----------------------------------------------------------------------------
\begin{figure*}
\centering
    %\hspace{-0.3in}
    \subfloat{\includegraphics[width=0.48\textwidth,angle=0]{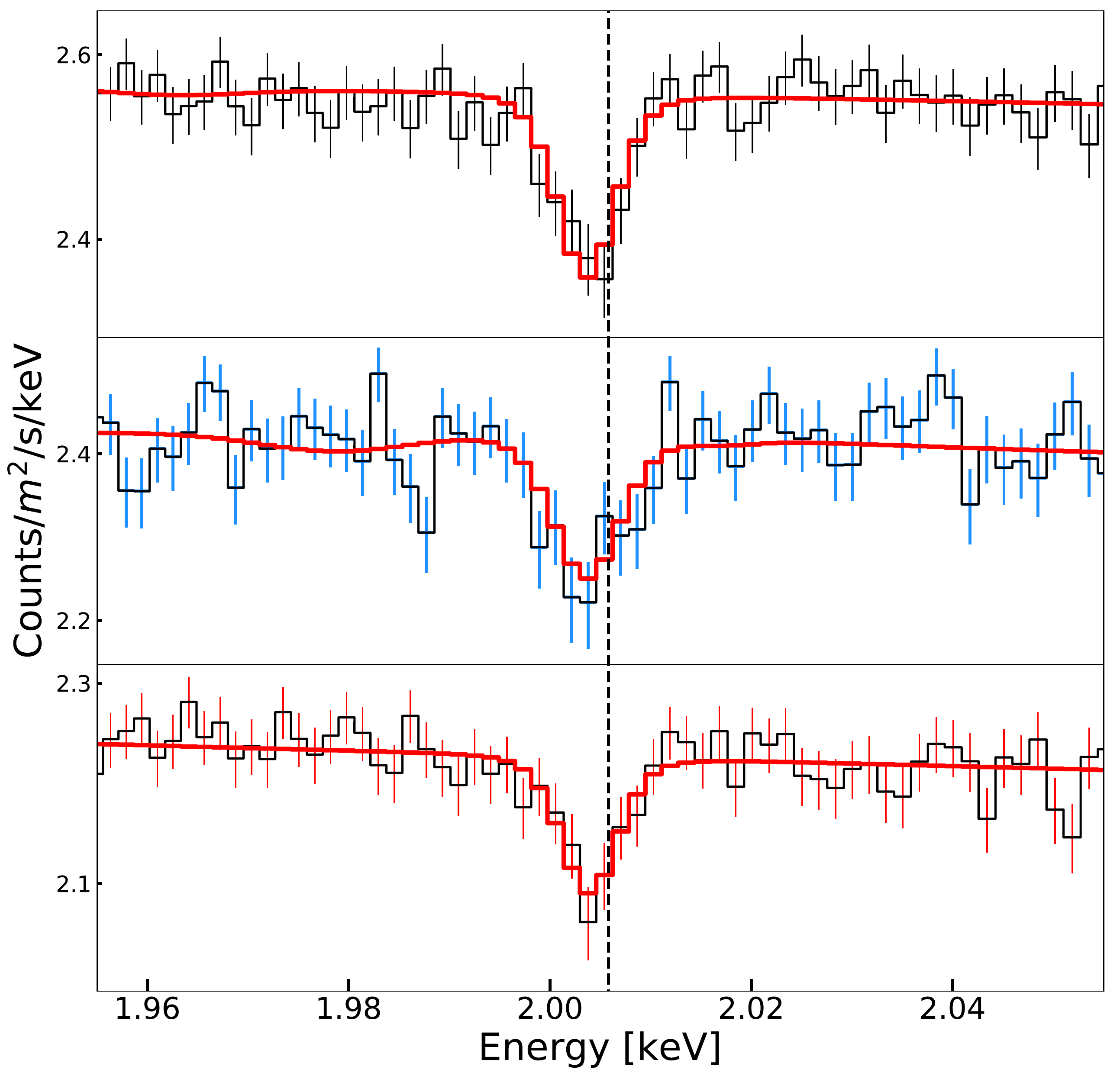}}
    %\hspace{-0.3in}
    \subfloat{\includegraphics[width=0.48\textwidth,angle=0]{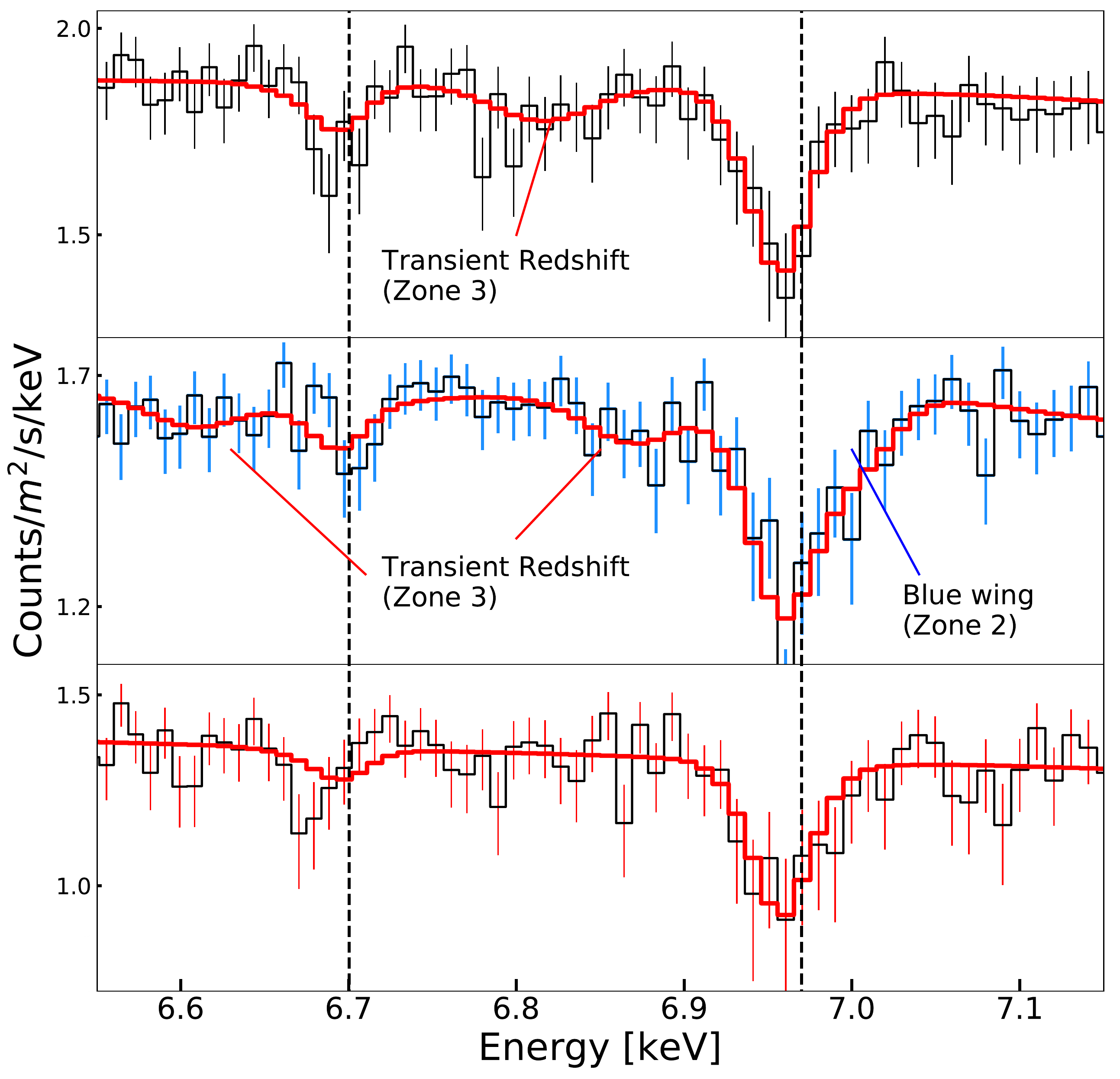}}
   %\vspace{-0.2in}

  \figcaption[t]{\footnotesize From top to bottom, the HETG spectra for Spectrum $\Gamma$, A, and B, with the Si XIV region (MEG) and Fe K band (HEG) plotted in the left and right panels, respectively. 
  The best-fit model listed in Table \ref{tab:fit} plotted in red. Black dotted vertical lines plot the rest energies for Si XIV$\alpha$ (2.0055 keV), Fe XXV$\alpha$ (6.70040 keV), and Fe XXVI$\alpha$ (6.96607 keV).  (The weighted average of the H-like spin-orbit doublets is quoted and plotted.)  The bulk of Si XIV and Fe XXVI, including line centers, display clear and consistent redshifts ranging from 230 to 290 $\text{km}$ $\text{s}^{-1}$. A blue-wing can be seen in the Fe XXV and Fe XXVI lines in Spectrum A (middle-right panel). 
   \label{fig:redshift} }
\end{figure*}
%-----------------------------------------------------------------------------

An important consideration in this analysis was the composition of the absorber. Optical studies of 4U 1916$-$053 strongly indicate that the companion in the system is likely a Helium-rich white dwarf \citep{Joss1978,Nelemans2006}; therefore, the disk in 4U~1916$-$053 is likely also Helium-rich. To first order, only the Hydrogen and Helium content is altered when a progenitor star evolves into Helium-rich white dwarf. Assuming a progenitor with standard solar abundances, the metal-to-helium ion abundances by number are $\sim 3 \times$ the metal-to-Hydrogen ion solar abundances. PION assumes a Hydrogen-rich absorber. Although individual abundances can be set for every ion (up to z = 30), it is unclear how setting an arbitrarily low Hydrogen abundance relative to Helium affects the calculation of the ionization balance or the normalization of the re-emission component. Instead, we adopted a more conservative approach wherein we increased the metal abundances relative to Hydrogen by a factor of 3, therefore treating each Hydrogen ion as a proxy Helium ion. This ensures that we obtain the correct absorbing columns while ensuring that the ionization balance calculations work as intended. Although this approximation underestimates the ion to electron ratio by a factor of two, the electron density mostly affects plasmas where collisional ionization is prominent, while it is the ion number density that sets the photoionization balance. 

All spectra were initially fit with a two absorption zone model: the first (Zone 1) accounting for the bulk of the absorption from the atmosphere, plus an additional blue-shifted absorber (Zone 2) to model any residual absorption owing to a possible disk wind. Dynamically broadened re-emission was added to models when required by the data, assuming a covering factor of $\Omega/4\pi=0.4$. This choice of covering factor is informed by the upper limit of $\Omega/4\pi \leq 0.5$ (based on a spherical shell where only half is not obscured by the accretion disk itself) and by a minimum covering factor given by the inclination of the source.  As will be clear in the following section, this choice of covering was ultimately inconsequential. In addition, if a second component was found not to be statistically significant for a particular spectrum, it was removed and the spectrum was fit again with a single absorber. In the case of Spectrum $\Gamma$ and A, we included an additional PION layer (Zone 3) to test the significance of the weak features at $\sim 6.8$ and $\sim 6.9$ keV using a highly redshifted and broad absorber.

\subsection{Fits}\label{sec:fits}

Parameters for the best-fit models to Spectrum A, B, and $\Gamma$ are listed in Table \ref{tab:fit}. 
For simplicity, we only list the model that best describes each spectrum. In all observations, the bulk of the absorption occurs in Zone 1 in the form of a disk atmosphere: a (nearly) static absorber with consistent properties, such as high absorbing columns (${N}_{He} \sim 40$ $\times$ ${10}^{22}$ ${\text{cm}}^{-2}$) and ionization (log $\xi \sim 4.4$), across all observations. Absorption owing to a static disk atmosphere of this kind is not only common, but expected among near edge-on ultra-compact systems. Of the three spectra, only Spectrum A required an additional, blue-shifted absorber (Zone 2) in order to capture both the blue-wing in the Fe XXVI line profile and the shift in Si XIV simultaneously. We discuss this possible wind in section \ref{sec:wind}.

Strikingly, the disk atmospheres in all three spectra show a consistent and unambiguous redshift of $\sim 250$ $\text{km}$ $\text{s}^{-1}$ that is evident and statistically significant (see Section \ref{sec:redshift}) in both the Fe XXVI and Si XIV lines (the two most prominent lines), and in agreement with other absorption features. The redshift is not a product of fitting a skewed line with a red wing in what otherwise would be a static absorber. As can be seen in Figure \ref{fig:redshift}, the line centers of both Fe and Si lines are redshifted to the best-fit value and the bulk of their equivalent widths are below their rest energies (vertical dashed lines). This redshift is also present regardless of whether one or two absorbers are used to model each spectrum. We discuss the origin of the redshift in Section \ref{sec:redshift}.

\subsection{The Persistent Redshift}\label{sec:redshift}

The detection of ionized absorption lines at or near their rest-energies in near edge-on sources is clear evidence of a static atmosphere above the disk surface, as these are nearly ubiquitous at high-inclinations and absent at moderate to low inclination sources. 
As noted in Section \ref{sec:fits}, the atmospheric absorption found in all the Chandra/HETG spectra of 4U 1916$-$053 (which encompasses the majority of the absorption) has a consistent redshift ranging from 230 to 290 $\text{km}$ $\text{s}^{-1}$. The presence of this shift is notable, as it is unusually high compared to the expected relative radial velocity (and dispersion) in the galaxy and could be evidence of a kick. If the radial velocity of system is small, however, the shift could instead originate in an inflow or a gravitationally redshifted atmosphere. In this section we analyze each of these possibilities.

First, we established the significance of the redshift by performing a simultaneous fit of the three spectra.
We performed three variations of the same test, where we tracked the increase in ${\chi}^{2}$ between the best-fit redshifted atmosphere model and the best-fit atmosphere model with velocities fixed at zero. In the first test, the absorber used to fit the three spectra shared the same column, ionization, and velocity. In the second test, only the velocity was linked between the three spectra, while all parameters were fit independently in the third test. We obtained $\Delta{\chi}^{2}$ values of 36, 37, and 38 for 3, 7, and 9 free parameters, respectively. This corresponds to a significance for the first test above 5$\sigma$ and above 4$\sigma$ for the second and third tests. The significance of the redshift is high despite using only one absorption zone with no re-emission to model all three spectra. Finally, we added the three spectra into a single combined A + B + $\Gamma$ spectrum and fit the time-averaged absorption. We obtained an average redshift of ${260}^{-80}_{+80}$ $\text{km}$ $\text{s}^{-1}$ with lower limits of 150 ($2\sigma$),  120 ($3\sigma$), 90 ($4\sigma$), and 50 ($5\sigma$) $\text{km}$ $\text{s}^{-1}$. These results are summarized in Table \ref{tab:reds}.

An overview of the \emph{Chandra X--Ray Observatory's} performance \citep{Marshall2004} reports average systematic errors on the absolute wavelength calibration for both the MEG and HEG of less than 100 $\text{km}$ $\text{s}^{-1}$. Systematic errors in the absolute wavelength calibration of the HETG arise from reconstructing the wavelength grid relative to the location zeroth order of the spectrum. This absolute wavelength calibration error, however, corresponds to the error in \emph{individual} lines in \emph{individual} spectra, as opposed to the exercise of determining the shift of an absorber by fitting \emph{multiple lines}. The aforementioned study demonstrates how the \emph{average} error in the shift of the \emph{emitter} is considerably lower once multiple lines are considered. Given that our 3 spectra were produced by combining 11 total observations, each with their own independent spectral extraction (and therefore independent wavelength error) yet still produce a consistent redshift suggests that systematic errors in the energy grid of the HETG are not significant. Much smaller uncertainties on velocities, such as an analysis on Capella which achieved errors at the $90\%$ level of $\sim 30$ $\text{km}$ $\text{s}^{-1}$ \citep{Ishibashi2006}, are commonly accepted.

%-----------------------------------------------------------------------------
\begin{figure}
\centering
    \subfloat{\includegraphics[width=0.48\textwidth,angle=0]{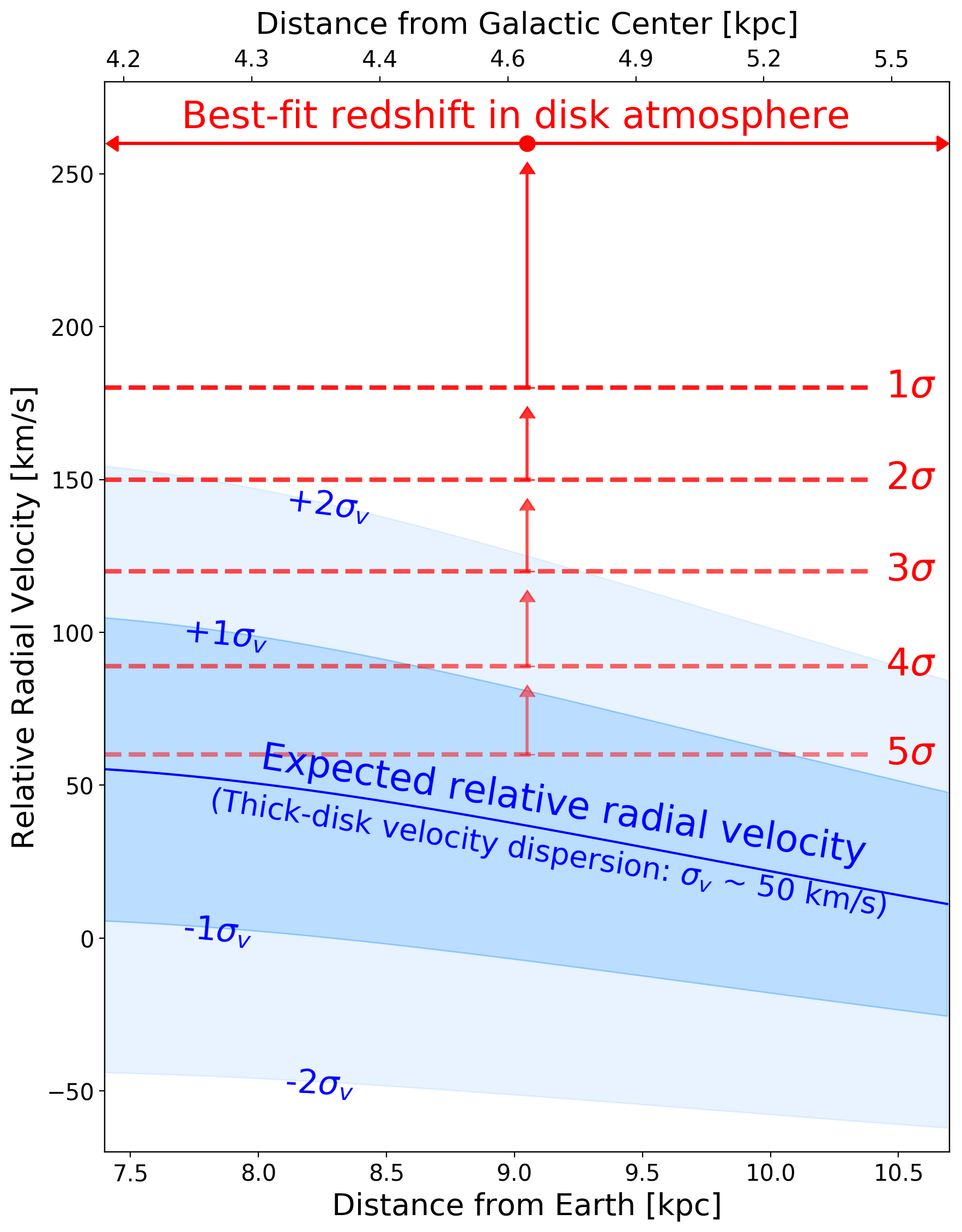}}
   %\vspace{-0.2in}
  \figcaption[t]{\footnotesize The predicted relative radial velocity of 4U 1916$-$053 as a function of its distance from the galactic center, based on its distance from earth ($9.0 \pm 1.3$ kpc) and galactic coordinates. The mean velocity is plotted in blue using a thick-disk rotational lag (relative to the thin-disk) of $\sim 50$ $\text{km}$ $\text{s}^{-1}$, while the the shaded regions represent integer multiples of the thick-disk velocity dispersion ($\sim 50$ $\text{km}$ $\text{s}^{-1}$). The best-fit disk atmosphere redshift for the combined A $+$ B $+$ $\Gamma$ spectrum is plotted in red, with the 1$\sigma$ to 5$\sigma$ confidence regions plotted as dashed lines. 
  The small overlap between $2\sigma$ confidence regions below $\sim$7.7 kpc is eliminated once uncertainties in the distance are folded in. The observed redshift cannot easily be attributed to the kinematics of the thick disk. 
   \label{fig:RV}}
\end{figure}
%-----------------------------------------------------------------------------

%-----------------------------------------------------------------------------
\begin{figure*}
\centering
  \subfloat{\includegraphics[width=0.48\textwidth,angle=0]{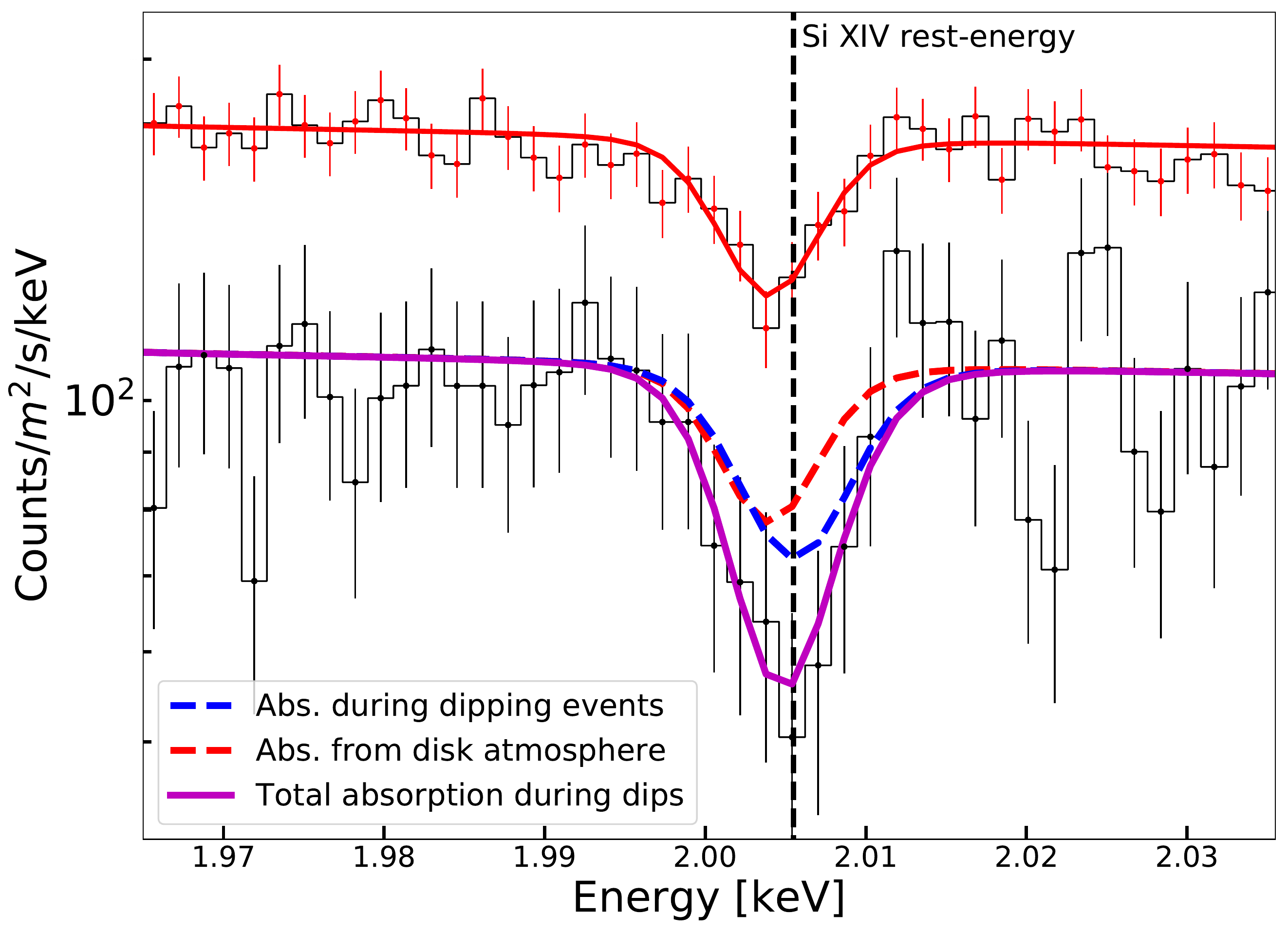}}
    \hspace{0.1in}
    \subfloat{\includegraphics[width=0.48\textwidth,angle=0]{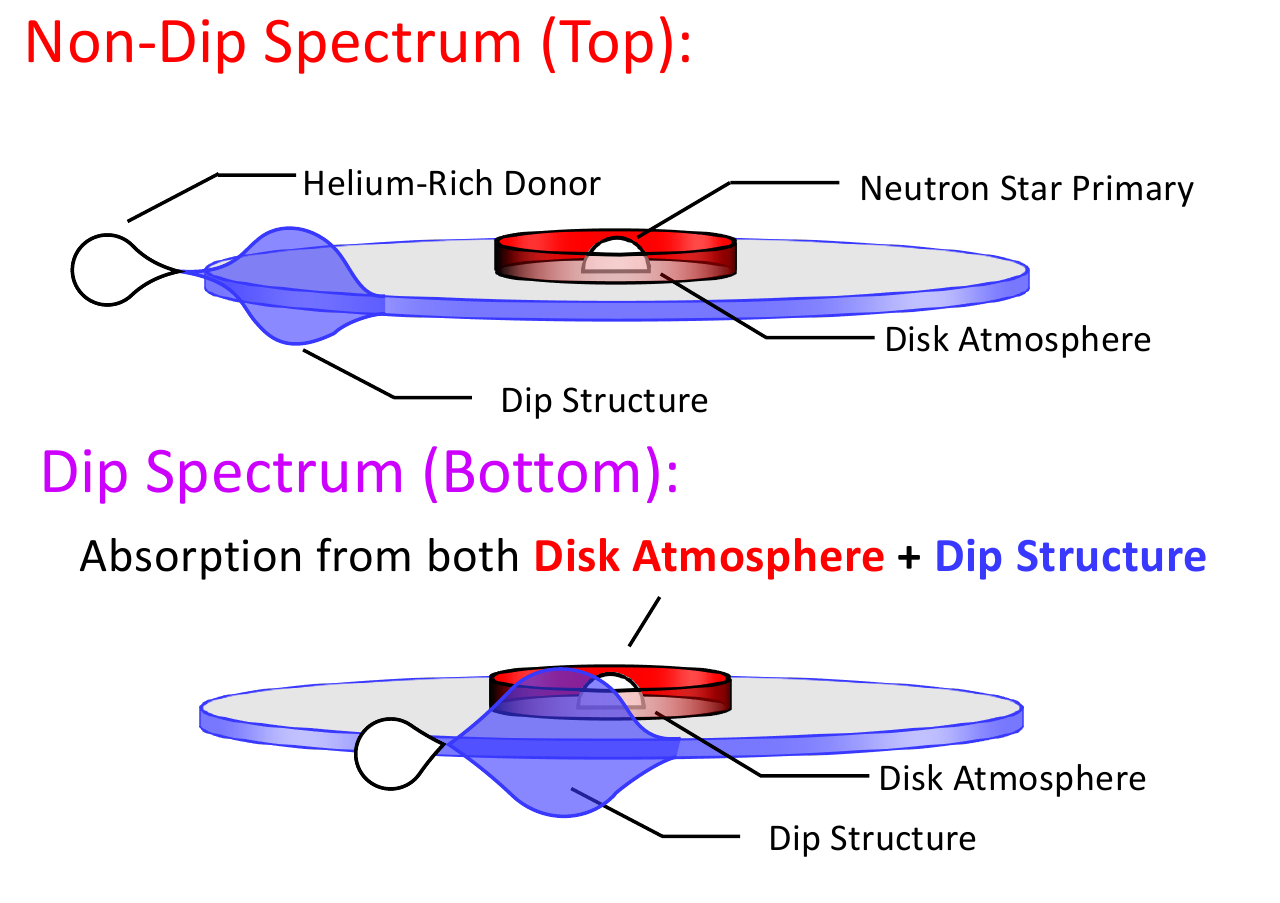}}

  \figcaption[t]{\footnotesize Spectrum B dip (bottom, black) and non-dip (top, red) MEG spectra in the Si XIV region. The redshifted atmospheric absorption during non-dip spectra does not capture the Si XIV line during dip events, as well as other lower-ionization lines below 2 keV. The excess absorption is best-fit with a slightly \emph{blue-shifted} absorber (${v}_{abs} = {-80} \pm 110$ $\text{km}$ $\text{s}^{-1}$), where the $2\sigma$ confidence regions of this velocity shift in the outer disk (${v}_{dip} < 90$ $\text{km}$ $\text{s}^{-1}$) and the redshift in the disk atmosphere (${v}_{atm.} > 150$ $\text{km}$ $\text{s}^{-1}$) do not overlap. This suggests that the systematic radial velocity of the source is small and not responsible for the persistent redshift observed in the disk atmosphere.
   \label{fig:Dips}}
\end{figure*}
%-----------------------------------------------------------------------------

In galactic coordinates, 4U 1916$-$053 is located -8.5 degrees away from the galactic plane. With distance estimates ranging from 7 to 11 kpc \citep{Smale1988, Yoshida1995}, it lies at $\sim 1.5$ kpc above the disk of the milky way, most likely making it a member of the thick disk of the Milky Way. Figure \ref{fig:RV} shows the expected systematic velocity of the source for a range of distances, where we adopted a thick-disk rotational lag (relative to the thin-disk, $\sim 220$ $\text{km}$ $\text{s}^{-1}$) of $\sim 50$ $\text{km}$ $\text{s}^{-1}$ and a thick-disk velocity dispersion of $\sim 50$ $\text{km}$ $\text{s}^{-1}$ \citep{Pasetto2012}. The mean relative velocity is plotted in blue and the shaded regions represent 1$\times$ and 2$\times$ the local velocity dispersion. Based on the recent distance estimate of $9.0 \pm 1.3$ kpc \citep{Galloway2008}, the expected velocity of the system is ${37} \pm 54$ $\text{km}$ $\text{s}^{-1}$ with a $2\sigma$ upper limit of 142 $\text{km}$ $\text{s}^{-1}$, below the $2\sigma$ lower limit on the redshift in the disk atmosphere of 150 $\text{km}$ $\text{s}^{-1}$. Even without constraints on the distance, there is little to no overlap between the $2\sigma$ regions in Figure \ref{fig:RV}.

Although the predicted galactic motions are small, a \emph{natal kick} may produce a radial velocity corresponding to the observed redshift. Indeed, pulsar kicks have been observed in the 100-1000 $\text{km}$ $\text{s}^{-1}$ range \citep[see][]{Atri2019}, making it difficult to rule-out radial motion. If the redshift is indeed the result of the radial velocity of the system, however, then all absorption features otherwise expected to be at rest velocities should display the same redshift.  In 4U 1916$-$053 and other dippers, particularly strong absorption lines are detected during dipping events, and typically associated with the outermost edge of the accretion disk \citep{DiazTrigo2006}.  Based on the orbital phase and kinematics of dipping events, the velocity shift of these excess absorption lines from the outer disk should be dominated by the systematic velocity of the source.  All dips were initially filtered-out from the spectra analyzed in this work. However, the 250~ks of new {\it Chandra} exposure makes it possible to extract sensitive spectra from the numerous dip events in Spectrum B. 

\begin{table*}[t]
\renewcommand{\arraystretch}{1.0}
\caption{Velocity Shift Comparisons}
\vspace{-1.0\baselineskip}
\begin{footnotesize}
\begin{center}
\begin{tabular*}{\textwidth}{l c@{\extracolsep{\fill}} c l c c c   c  }
\tableline
\tableline
\multicolumn{3}{l}{\textbf{Disk Atmosphere Redshift}} & \multicolumn{4}{l}{\textbf{Relative Radial Velocity}}\\
\multicolumn{3}{l}{Combined A$+$B$+\Gamma$ Spectrum}& & & Expected Galactic Motion & Dip Absorption & Posterior Probability\\
&&($\text{km}$ $\text{s}^{-1}$)&&&($\text{km}$ $\text{s}^{-1}$)&($\text{km}$ $\text{s}^{-1}$)&($\text{km}$ $\text{s}^{-1}$)\\
\\ [-3.0ex]
\tableline 
\tableline 
\\ [-3.0ex]
Best-fit value && ${260}_{-80}^{+80}$ &\multicolumn{2}{l}{Best-fit or mean value:} & ${40} \pm 50$ & ${-80} \pm 110$ &${5} \pm 42$\\
Lower Limits:&$2\sigma$ & $>150$  &Upper Limits:& $2\sigma$ &  $<$\textbf{140}  & $<$\textbf{90}  & 85 \\
&$3\sigma$ &$>120$ &&$3\sigma$ & 190 & 170 & $<$\textbf{130}\\
&$4\sigma$&$>90$ &&$4\sigma$ & ... & ... & ...\\
&$5\sigma$&$>60$ &&$5\sigma$\\

\\ [-3.0ex]
\tableline 
\tableline

\end{tabular*}
\vspace*{-1.0\baselineskip}~\\ \end{center} 
\tablecomments{Comparison between the redshift measured in the disk atmosphere (left) and the radial velocity of the system (right). Best-fit or mean expected values are quoted with their $1\sigma$ confidence regions. Lower limits to the redshift in the disk atmosphere and upper limits to the systematic radial velocity of the system are quoted based on their $2\sigma$, $3\sigma$, $4\sigma$, and $5\sigma$ errors. Values in bold correspond to the closest integer significance at which the redshift in the disk atmosphere does not match the radial velocity. The posterior probability resulting from the thick-disk kinematics and the measured excess absorption in the outer disk yields a near $3\sigma$ difference from the shift in the disk atmosphere.}

\end{footnotesize}
\label{tab:reds}
\end{table*}

If the inner disk atmosphere is truly static, and if the dips truly originate in the outermost disk, then dip spectra should contain absorption lines from the inner disk atmosphere {\em and} the outermost disk.  We therefore modeled the dip spectrum using the same best-fit model for Spectrum B listed in Table \ref{tab:fit}, plus an additional absorber.  Due to the modest S/N in the Fe~K region of the dip spectrum, we focused only on the MEG spectrum between 1.5 and 2.5 keV. Our resulting fits suggested a relatively low ionization log ${\xi} \sim 2.2$, a helium column of ${N}_{He} \sim {10}^{22}$ $\text{cm}^{-2}$, and a small turbulent velocity of ${v}_{turb}\sim 50$ $\text{km}$ $\text{s}^{-1}$.

Figure \ref{fig:Dips} shows the Si XIV line for both dip (black) and non-dip spectra (red) with their corresponding best-fit models. The rest-energy of the Si XIV line is plotted as the vertical dashed line, and the contribution of the redshifted atmosphere is plotted in red for both spectra. It is clear that there is significant absorption in the dip spectrum that is not being captured by the absorber used to model the atmosphere in the non-dip spectrum. Visually, the excess absorption does not share the same velocity shift as the line originating in the atmosphere (red), with most of the excess dip absorption contributing at slightly higher energies, consistent with no velocity shift.  The combined line profile itself is not quite centered at the rest-energy of Si XIV.  This is confirmed by spectral fitting, where the excess (plotted in blue) is best-fit with a slightly \emph{blue-shifted} ionized absorber ${v}_{dip} = {-80} \pm 110$ $\text{km}$ $\text{s}^{-1}$ (with a $2\sigma$ upper limit of 90 $\text{km}$ $\text{s}^{-1}$). Despite the lower S/N of the dip spectrum, we can rule out the possibility of both the atmosphere and dips absorption sharing the same redshift at a modest confidence level of 2.5$\sigma$ by fitting the strongest lines in the 1.5 to 2.5 keV range. Indeed, the posterior probability resulting from combining the thick disk kinematics (the prior) with this outer disk shift suggests a near zero velocity shift and a 3$\sigma$ upper limit of $\sim130$ $\text{km}$ $\text{s}^{-1}$.

The lack of a \emph{positive} velocity shift in the outer disk suggests that the systematic velocity of 4U 1916$-$053 (its projected kick velocity, if any) is significantly lower than the $\sim$260 $\text{km}$ $\text{s}^{-1}$ redshift observed in the atmosphere, and formally consistent with zero. This analysis strongly suggests that the persistent redshift seen in all spectra is associated with an inner disk atmosphere.

A failed wind or inflow could, in principle, be responsible for the observed redshift.  The most direct evidence of such a flow would be the presence of inverse P-Cygni profiles.  These are missing from our data, yet ruling out an inflow in this fashion likely requires much higher quality spectra. Still, an inflow of this sort seems unlikely. First, an inflow would require some mechanism that removes enough angular momentum from the atmosphere in a much lower density environment compared to those found in the disk. This mechanism would have to produce an inflow with a very low covering factor, as we do not observe ionized absorption lines, much less redshifted lines, in sources viewed at lower inclinations. This means the inflow, which must originate at relatively large radii, would have a small vertical extent.  The presence of X-ray bursts suggest that the magnetic field of the neutron star is likely not strong enough to induce a collimated inflow. 

The specific properties of the absorber also render unlikely an inflow. In the almost 300ks of new and archival Chandra/HETG exposures of 4U 1916$-$053, the source displays a range of spectral states over a period of $\sim$14 years. Despite the large temporal separation, the observed redshift appears to be constant for all observations even as source undergoes noticeable changes in accretion state. This would suggest that whatever physical mechanism was responsible for removing the angular momentum from the atmosphere would necessarily produce the same redshift despite different physical conditions within the disk. 

We find that the most likely explanation for this redshift is that of a static atmosphere, located close enough to the compact object to be subject to a $\sim$260 $\text{km}$ $\text{s}^{-1}$ gravitational redshift. Unlike a putative inflow, a gravitationally redshifted disk atmosphere (or, GRDA for short) would be more consistent with the static atmospheres we expect to see in these sources and with the behavior of the absorption in this source in particular. This explanation only invokes general relativity in the weak limit, requiring the absorber to be located at around $\sim 1200$ $GM/{c}^{2}$, or $\sim 1800$ $(\frac{{M}_{NS}}{{M}_{\odot}})$ km. 

Table \ref{tab:atm} lists the properties of the disk atmosphere found in our spectra, using three different methods for measuring the radius from the compact object to absorber. The first is the upper limit on the photoionization radius, ${R} < {R}_{upper} \equiv L/{N}_{He}\xi$, where ${R} = f\cdot{R}_{upper}$ and $f$ is the filling factor. We assumed a filling factor of unity for simplicity. The second is the radius given by the dynamical broadening of the re-emission from the gas, $R = G M/{\sigma}_{emis}^{2}$, assuming the velocity broadening (${\sigma}_{emis}^{2}$) probes the local Keplerian velocity.
In the case of a GRDA, the measured redshift can be used in order to directly measure this radius in units of $GM/{c}^{2}$, assuming the redshift is entirely due to gravity. The radius, $R = {R}_{z}$, is also included in Table \ref{tab:atm}. 

\begin{table*}[t]
\renewcommand{\arraystretch}{1.2}
\caption{Disk Atmosphere Geometry and Physical Properties}
\vspace{-1.0\baselineskip}
\begin{footnotesize}

\begin{center}
\begin{tabular*}{\textwidth}{l   l @{\extracolsep{\fill}}  c c  c  cc c c}
\tableline
\tableline

Spectrum &  & \multicolumn{2}{c}{Radius} & ${n}_{He}$ & $f$ & ${z}_{Grav.}$  & ${v}_{corr.}$ & ${v}_{corr,2{M}_{\odot}}$\\
&  & (${10}^{8}$ cm) & (${GM}/{c}^{2}$) & (${10}^{14}$ ${\text{cm}}^{-3}$)& &($\text{km}$ $\text{s}^{-1}$) & ($\text{km}$ $\text{s}^{-1}$)  & ($\text{km}$ $\text{s}^{-1}$) \\

\tableline 
Spectrum A&
${R}_{Upper}$   & 
${5.4}^{+2.0}_{\ddagger}$ &
${3600}^{+1300}_{\ddagger}$ &
${9.3}^{\ddagger}_{-4.5}$ &
${1.0}^{\dagger}$&
${70}^{\ddagger}_{-30}$ &
${160}^{+230}_{-220}$ & ${90}\pm 230$\\

&${R}_{Orbital}$   & 
${4.65}^{+4.7}_{-4.1}$ &
${3100}^{+3200}_{-2800}$ &
${13}^{+13}_{-10}$ &
${0.9}^{\ddagger}_{-0.8}$ &
${100}^{+90}_{-100}$ &
${120}^{+500}_{-240}$ &$-$ \\

&${R}_{z}$   & 
${2.0} \pm 1.7$ &
${1300} \pm 1200$ &
${70}^{+130}_{-50}$ &
${0.4}^{\ddagger}_{-0.4}$ &
${260}^{\dagger}$ &
${0}^{\dagger}$ &$-$ \\

\\ [-3.0ex]
\tableline 
Spectrum B&
${R}_{Upper}$   & 
${5.4}^{+1.8}_{-4.4}$ &
${3700}^{+1200}_{-3000}$ &
${6.0}^{+5.2}_{-4.9}$ &
${1.0}^{\dagger}$&
${80}^{+70}_{-30}$ &
${150}^{+160}_{-140}$ &${60}^{+200}_{-140}$  \\

&${R}_{Orbital}$   & 
$-$ &
$-$ &
$-$ &
$-$ &
$-$ &
$-$ &$-$\\

&${R}_{z}$   & 
${2.0}^{+1.1}_{-1.2}$ &
${1300} \pm 800$ &
${50}^{+30}_{-40}$ &
${0.36}^{+0.36}_{-0.25}$ &
${228}^{\dagger}$ &
${0}^{\dagger}$  &$-$\\
\\ [-3.0ex]
\tableline 
Spectrum $\Gamma$&

${R}_{Upper}$   & 
${17}^{+3}_{-11}$ &
${11000}^{+2000}_{-7000}$ &
${4.0}^{+2.6}_{-1.9}$ &
${1.0}^{\dagger}$&
${26}^{+16}_{-5}$ &
${260}^{+170}_{-160}$ & ${230}^{+170}_{-160}$\\

&${R}_{Orbital}$   & 
${1.5}^{+0.9}_{-1.0}$ &
${1000} \pm 600$ &
${480}^{+300}_{-360}$ &
${0.09}^{+0.08}_{-0.06}$ &
${290}^{+180}_{-170}$ &
${0}^{+240}_{-230}$ & $-$\\

&${R}_{z}$   & 
${1.6} \pm 0.9$ &
${1100} \pm 600 $ &
${500} \pm 300 $ &
${0.09}^{+0.08}_{-0.06}$ & 
${285}^{\dagger}$ &
${0}^{\dagger}$ & $-$ \\

\\ [-3.0ex]
\tableline 
\tableline

\end{tabular*}
\vspace*{-1.0\baselineskip}~\\ \end{center} 
\tablecomments{Physical properties of the disk atmosphere of Spectrum A, B, and $\Gamma$ as a function of the upper limit of on the photoionization radius (${R}_{Upper} =f\cdot L/{N}_{He}\xi$), the radius given by the dynamical broadening of the re-emission (${R}_{Orbital}$), and the radius given by the observed redshift (${R}_{z}$). 
A neutron star mass of $1{M}_{\odot}$ was assumed to calculate some values. When using ${R}_{Upper}$, only ${z}_{Grav}$ and ${v}_{corr.}$ depend on mass. Values for ${v}_{corr.}$ assuming at $2{M}_{\odot}$ are listed in the last column.  A Filling factor of unity is assumed using ${R}_{Upper}$. In the case of both ${R}_{Orbital}$ and ${R}_{z}$, only the density (${n}_{He}$) and filling factor ($f$) depend on mass. For $2{M}_{\odot}$ neutron star, the density decreases by a factor of 4 and the filling factor increases by a factor of 2.} 

\end{footnotesize}
\label{tab:atm}
\end{table*}

For simplicity, the gravitational redshift (${z}_{Grav}$) and the resulting corrected velocity (${v}_{corr}$) values at $R = {R}_{Upper}$ listed in Table \ref{tab:atm} were calculated assuming a neutron star mass of $1{M}_{\odot}$ (values for ${v}_{corr}$ for $2{M}_{\odot}$ were also included). Being an upper limit on the distance of the absorber from the central engine, ${R}_{Upper}$ places a lower limit on ${z}_{Grav}$ which only increases with neutron star mass. In the case of Spectrum A and B, ${R}_{Upper}$ suggests that the disk atmosphere must be gravitationally redshifted by at least $\sim$70 $\text{km}$ $\text{s}^{-1}$, which increases to $\sim$150 $\text{km}$ $\text{s}^{-1}$ for a mass of $2{M}_{\odot}$. In the case of Spectrum $\Gamma$, this lower limit is significantly smaller.  Uncertainties on the distance and SED shape make ${R}_{Upper}$ an imperfect upper limit.  It is notable that the radius given by the broadening of the re-emission, $R = {R}_{Orbital}$, is consistent with the radii determined by associating the redshift with a gravitational shift. However, this velocity broadening value is poorly constrained and assumes an semi--arbitrary choice of emission covering factor, and therefore we advise caution when interpreting this result.  Ultimately, the picture of a static, photoionized disk atmosphere deep in the gravitational potential is self-consistent.  Please see Table 3 for all of the relevant measurements, and velocities corrected to the assumed rest frame of the absorption.

If the observed redshift is entirely gravitational, it is possible to constrain other aspects of the disk atmosphere by insisting on the radii being equal.
For instance, the filling factors, $f$, for Spectrum A and B range from $\sim$ 0.3 to $\sim$0.8 for neutron star masses of 1 and 2 ${M}_{\odot}$, respectively. This suggests a low degree of clumping within the atmosphere; in this respect, the atmosphere may differ substantially from winds that are observed downstream in this and other sources.  At 1 ${M}_{\odot}$, the atmosphere in Spectrum $\Gamma$ would be significantly more clumpy by comparison, although this low filling factor of $f\sim0.1$ is far from unphysical.  It is also possible to derive gas densities in the atmosphere; these range from 15 $<$ log ${n}_{He} <$  16.7. This range of gas densities are comparable to those inferred in studies of disk winds in accreting black holes \citep[e.g.,][]{Miller2016, Trueba2019}. At the higher end, these values are comparable with densities expected within the disk (15 $<$ log ${n}_{e} <$  19, see \citealp{Garcia2016}). Finally, we note that when the radial location of the absorber is known, the local Keplerian velocity is defined; then, if the turbulent broadening can be measured via fits to multiple lines, the measured width of time-averaged absorption lines gives the size of the central engine.  Using these data, a value of $R \simeq {60}^{+20}_{-30}$  $GM/{c}^{2}$ results.  However, it is likely more appropriate to treat this as an upper limit. This technique is the subject of a forthcoming paper (Trueba et al.\ 2020, in preparation).

A disk atmosphere located at $\sim 1200~GM/c^{2}$ would likely require magnetic pressure support in order to produce the columns we observe along our line of sight, as any plausible gas temperatures produce far less thermal energy than required. 
Numerical models of MRI \citep[Magnetorotational instability, see][]{BalbusHawley} disks suggest than magnetic pressure support can not only dominate the upper layers of the disk, but can also result in a more extended vertical structure \citep{Blaes2006}, much like the absorbing atmosphere along our line of sight. As an order of magnitude comparison, we estimated the local magnetic field ($B \sim \sqrt{8\pi P}$) required to maintain hydrostatic equilibrium via $dP/dz \simeq sin~\theta\cdot GM\rho/{r}^{2}$, where $sin~\theta \simeq r/z$ and $dP/dz \sim P/z$, and therefore $P \simeq GM\rho/{r} \cdot {z}^{2}/{r}^{2}$. Using the largest densities in Table \ref{tab:atm} and a large $z/r~0.3$, we get a local magnetic field of $\sim 2~10^5~\text{G}$, which is still below the upper limit on the magnetic field given by \citet{SS73} of $\sim 5~10^5~\text{G}$ (equation 2.19).

\subsection{Redshifted Lines in Archival Data}\label{sec:others}

A large number of neutron star X-ray binaries are currently known, so it is unlikely that a gravitationally redshifted disk atmosphere would be unique to 4U 1916$-$053.  However, no redshifts like those now discovered in 4U 1916$-$053 have been reported in the literature.  In longer-period systems with larger disks, a longer path length through the extended disk atmosphere may simply wash-out the signature of the inner disk atmosphere.  Particularly amongst ultra-compact and short period dipping/eclipsing X-ray bursters, however, wherein the magnetic field of the neutron star is weak and the accretion disk is small, we should find evidence of redshifted atmospheres.  We examined the available Chandra/HETG archival data for sources that fit these criteria.  The same data reduction procedure as in section \ref{sec:data} was followed. 

Of the few sources in our sample, we found very similar redshifts of $\sim$300 $\text{km}$ $\text{s}^{-1}$ in the spectra of both XTE J1710$-$281 and AX J1745.6$-$2901.  As per the absorption seen in 4U 1916$-$053, it is the bulk of the line absorption that is redshifted, with multiple lines displaying the same redshift.  In the only HETG spectrum of AX J1745.6$-$2901, the redshift is  present in both He and H-like iron absorption lines, at a significance of $\sim 2.5\sigma$.  Unfortunately, the neutral ISM column is too high to confirm this redshift via Si XIV.  This is not the case for the brighter of the two archival spectra of XTE J1710$-$281, which instead showed multiple strong redshifted absorption lines between 1.0$-$2.6 keV as well as Fe~K. The $\sim$320 $\text{km}$ $\text{s}^{-1}$ redshift is unmistakable and significant at well over 5$\sigma$ (Trueba et al. 2020 in prep).  

The Chandra observation of AX J1745.6$-$2901 was a conducted as part of a campaign involving both XMM-Newton and NuSTAR \citep[see][]{Ponti2018}. To date, a spectral analysis of this spectrum has not been published and no redshift has been reported. This is likely due to the relatively low S/N of the spectrum.  More surprising is the case of XTE J1710$-$281, where the highly significant redshift has also not been reported.  It is possible that carefully screening against dips and bursts, and our use of the relatively new ``optimal'' binning algorithm, has enabled us to extract more information from new and archival data.

We note that possible redshifts were also found in the spectra of other sources (e.g., EXO 0748$-$676), but these were either low S/N spectra or displayed complex absorption with multiple velocities.  The remaining spectra contained either no absorption lines or, in the case of sources with longer orbital periods, lines at rest velocities.  Based on the galactic coordinates of both XTE J1710$-$281 (356.3571 +06.9220, likely in the thick disk) and AX J1745.6$-$2901 (359.92030 -00.04204), a natal kick would likely be required in order to attribute the observed redshifts entirely to the relative radial velocities of either of these systems. Ultimately, robust radial velocity measurements are required in order to confirm the origin of these redshifts. However, these results are promising in that they are consistent with a physical connection between the processes that give rise to disk atmospheres with redshifted lines, be it gravitational redshift from an inner disk atmosphere or some persistent, semi--spherical inflow.

\subsection{Transient Absorption Features}\label{sec:features}

In the previous section, we mentioned evidence of transient absorption features in our spectra aside from the primary absorber, which we attribute to a redshifted inner disk atmosphere. The first is evidence of absorption between 6.8 and 6.9 keV in both Spectrum A and $\Gamma$, which we attribute to a much more highly redshifted disk atmosphere. The second is evidence of a disk wind in Spectrum A, which appears as a blue wing in the Fe XXVI absorption line at $\sim 6.97$ keV (as well as perhaps some complexity in the Si XIV line near $\sim 2$ keV). In this section, we will discuss alternative interpretations and fits to the data, their significance, and some physical implications of these results.

\subsubsection{The Transient Redshift}\label{sec:transient}

The putative highly redshifted component deserves additional scrutiny and caution.  It is present in two of the three spectra in question (only missing from the lowest S/N spectrum: Spectrum B), suggesting it may not simply be noise.  In addition, if we divide the best-fit column density by its $-1\sigma$ error (analogous to assessing the significance of a Gaussian line using its normalization and error) we can obtain a crude estimate of the significance for this component. In the case of Spectrum $\Gamma$, possible fits to this component displayed a strong degeneracy between ${N}_{He}$ and log ${\xi}$ for values of ${N}_{He}$ above $\sim 20\times {10}^{22} $ ${\text{cm}}^{-2}$, which is why we froze this parameter in the fit reported in Table \ref{tab:fit}. In order to determine the significance in this spectrum, we re-fit the data by thawing the ${N}_{He}$ parameter and then performing an error search with the log ${\xi}$ parameter frozen at its new best-fit value. Freezing log ${\xi}$ was necessary in this case: for lower ${N}_{He}$ values, the fit will yield a lower ionization in order to produce a comparable line depths, making this test meaningless. We obtained a column of ${34}^{+4}_{-12}$ (${10}^{22}$ ${\text{cm}}^{-2}$), which corresponds to a significance of $\sim 3\sigma$. In addition, we performed an f-test comparing our model to a single-zone model and also obtined a significance of $3\sigma$.

This test was not as straightforward in the case of Spectrum A. Although visually these features are similar in both spectra, the errors on the column for this component suggest a significance of barely above 1$\sigma$. This, in part, is due to the complexity of the model for Spectrum A. As previously mentioned, an F-test suggests a $3\sigma$ improvement over the single zone model. However, given the low significance suggested by the errors on the ${N}_{He}$ parameter, we advise caution when interpreting the significance of this component.
Notably, we have established that the bulk of the observed absorption originates in an atmosphere that is redshifted, possibly as a result of local gravitational redshift.  This component may simply be a different portion of this atmosphere arising at smaller radii.

Alternatively, these components could be evidence of a highly \emph{blueshifted} absorber, corresponding to gas outflowing at velocities near $\sim 0.03c$. In both Spectrum $\Gamma$ and Spectrum A, however, the most prominent feature motivating the analysis of this transient absorber is located near $\sim 6.8$ keV, which in the case of a \emph{blueshifted abosorber} would correspond to He-like Fe XXV absorption. The location of the corresponding blueshifted Fe XXVI line shows little evidence of an absorption line, although weak Fe XXVI absorption is plausible given the quality of the data. Prominent Fe XXV lines with little Fe XXVI absorption suggest a gas of fairly low-ionization which is often accompanied by absorption from He-like and H-like Ca, Ar, Si, and S (to name a few), none of which we find at or near the blueshifts suggested by the feature at 6.8 keV, making the high-blueshift interpretation unlikely.

This was confirmed by spectral fitting, as we were completely unable to fit a blueshifted component onto the feature near $\sim 6.8$ keV in Spectrum A using a photoionized absorption model. In the case of Spectrum $\Gamma$, however, the significance of a highly blueshifted absorber (${\chi}^2/dof = 487/485$) is comparable to that of the highly redshifted one (${\chi}^2/dof = 482/486$). If the feature at $\sim 6.8$ keV is real in both spectra, then it seems more likely for these absorbers to originate via similar physical mechanisms. This means that it is unlikely that we are detecting a dramatically blueshifted absorber in one observation and a dramatically redshifted one in another, especially when the main feature in both is located at essentially the same energy. Our fits to Spectrum A would then suggest that this is more likely an extremely redshifted inner disk atmosphere rather than a near-UFO. Given the quality of the data, however, the nature of this feature will remain unclear until further observation.

\subsubsection{Evidence of a Disk Wind}\label{sec:wind}

Previous studies on the best Chandra/HETG LMXB wind spectra have been able to model and identify multiple velocity components of the absorbing wind complex.
In cases such as 4U 1630$-$472 \citep{Miller2015,Trueba2019} and GRS 1915$+$105 \citep{Miller2016}, although the individual velocity components are all blended to produce multiple skewed line profiles, the resulting line spectrum displays different velocity shifts for lines corresponding the different ionization and column densities of the separate absorbers. In these cases, models with multiple absorbers are {\em required} to describe the data. 

The same analysis can be applied to 4U 1916$-$053 where, instead of a wind complex, the presence of both a static atmosphere and a weakly absorbing wind may require the use of multiple absorbers.
Of the three spectra, compelling evidence for a disk wind can only be found in Spectrum A. 
Although the shape of the Fe XXVI line in Spectrum B is similar to that in Spectrum A, a two component model resulted in no significant improvement to the fit because of the lower S/N of the spectrum, as well as the asymmetry in the line being less pronounced.

%-----------------------------------------------------------------------------
\begin{figure}
\centering

    \subfloat{\includegraphics[width=0.48\textwidth,angle=0]{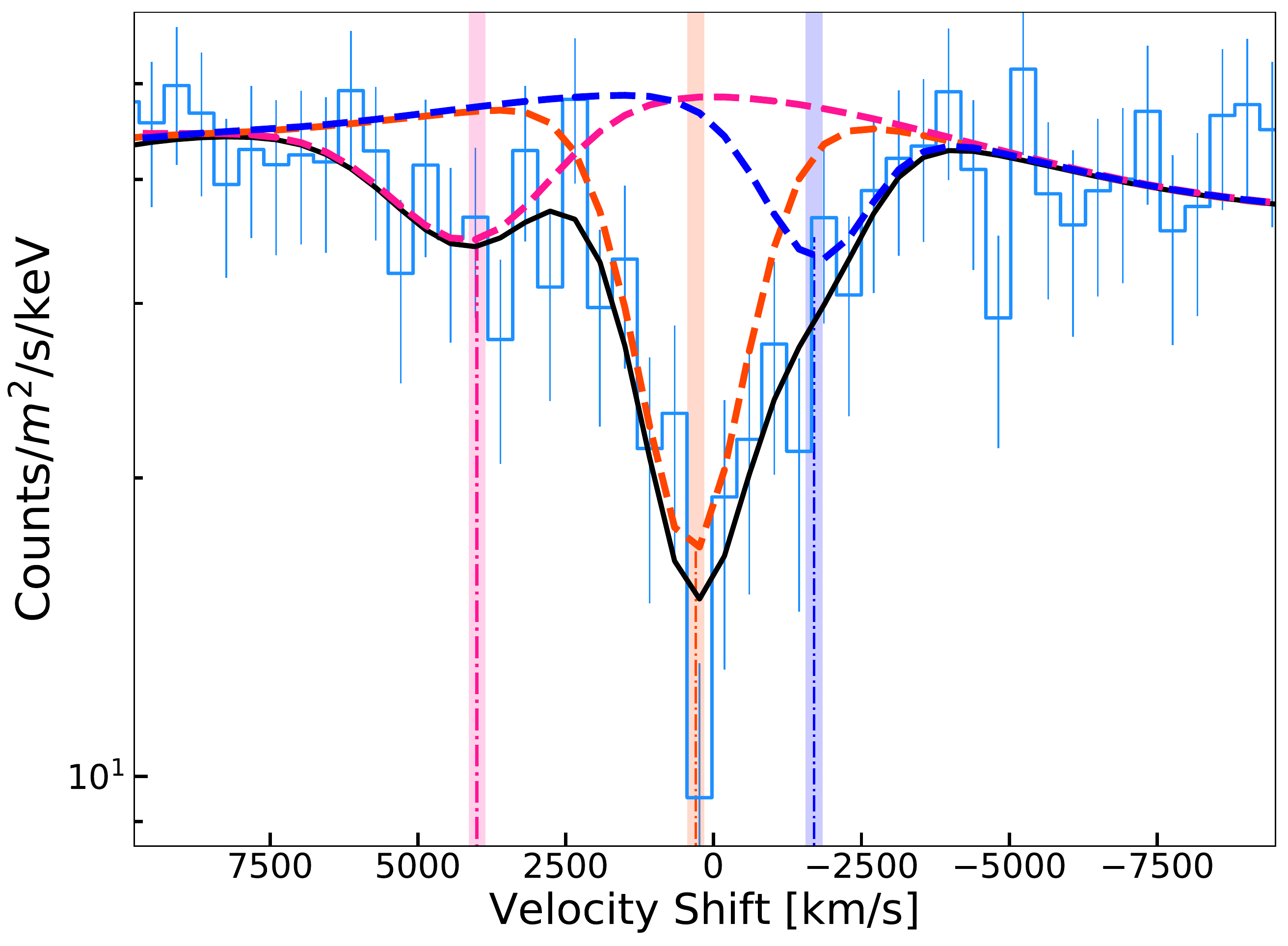}}

  \figcaption[t]{\footnotesize The combined first-order HEG portion of 4U 1916$-$053 from Spectrum A, with the best-fit 3-zone model. The combined Fe XXVI line profile is plotted in velocity space relative to the rest energy of Fe XXVI.
  The contribution of the putative highly redshifted absorber ($v \sim 4300$ $\text{km}$ $\text{s}^{-1}$, plotted in pink) can also be seen near $\sim 6.6$ and $\sim 6.8$ keV in Figure \ref{fig:redshift}.  A blue-shifted disk wind component ($v \sim -1700$ $\text{km}$ $\text{s}^{-1}$, plotted in blue) is required to fit the extended blue wing of the line.  Separately, the bulk of the composite line requires modestly redshifted Si and Fe absorption (plotted in red) owing to the disk atmosphere. In this model, the Si XIV line is captured by a single absorber (Zone 1, the disk atmosphere).
   \label{fig:wind} }
\end{figure}
%-----------------------------------------------------------------------------

%-----------------------------------------------------------------------------
\begin{figure}
\centering
\vspace{0.1in}
    \subfloat{\includegraphics[width=0.48\textwidth,angle=0]{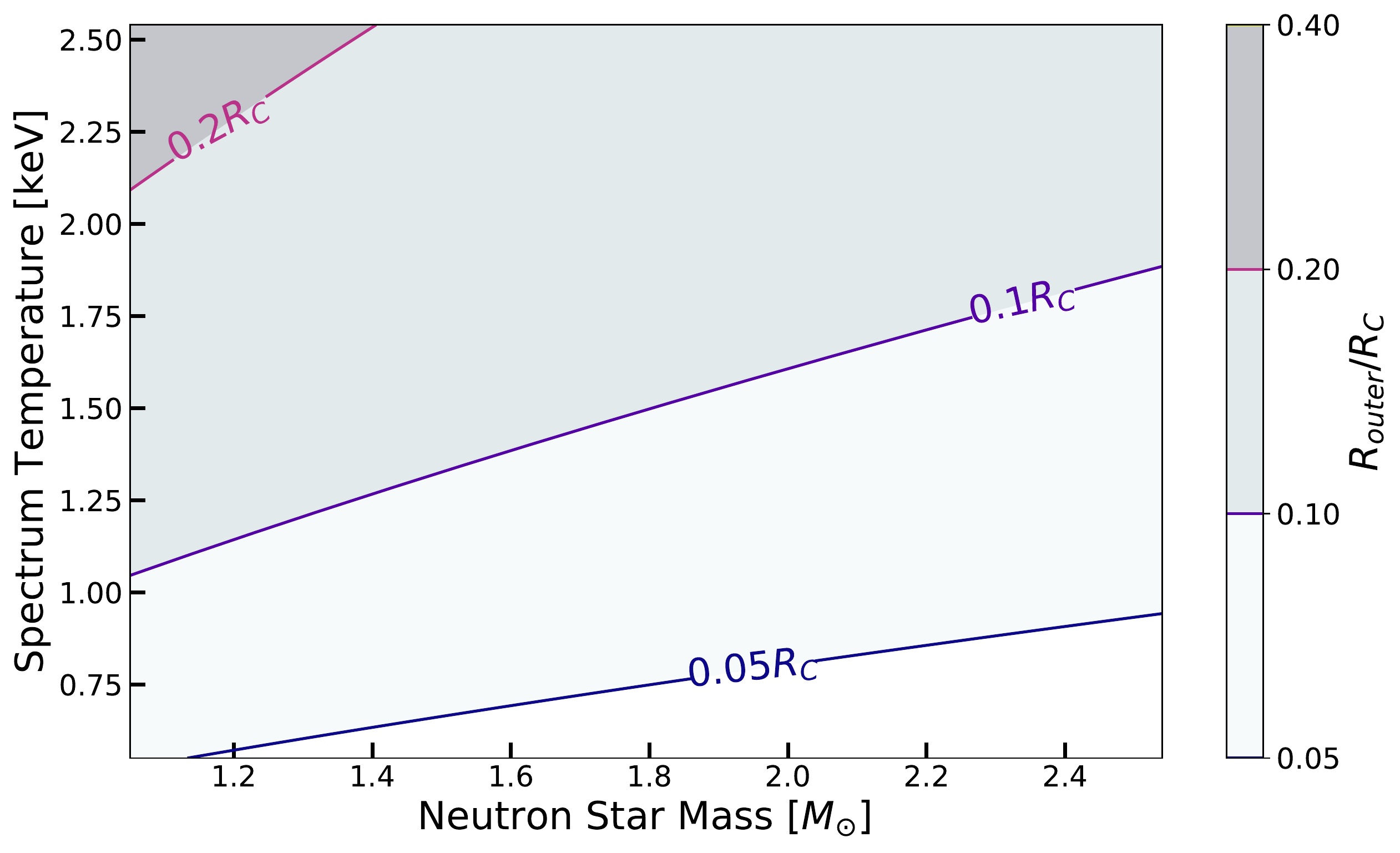}}
   %\vspace{-0.2in}
  \figcaption[t]{\footnotesize Ratio of the disk outer-radius (based on the orbit of the donor star, for an orbital period of 50 minutes) over the Compton radius, as function of neutron star mass and the temperature of the comptonizing spectrum.  Within the plausible range of temperatures (based on fitting the disk and neutron star temperatures) and neutron star mass, this outer radius never exceeds $0.2\times$ the Compton radius. 
   \label{fig:Rorb}}
\end{figure}
%-----------------------------------------------------------------------------

The three component model of spectrum A (including a wind; see Table \ref{tab:fit}) yields a ${\chi}^2/dof = 536/488 = 1.10$,
a modest but non-trivial improvement of $\Delta{\chi}^2 = 23$ over the best-fit single-component model (${\chi}^2/dof = 559/495 = 1.14$). 
An F-test quantifies this as a $3\sigma$ improvement over the one component model. 
Visually, the one component model is unable to capture shifts and depths of both the Si XIV and Fe XXVI simultaneously.
Since the redshift of the single absorber model is lower than that in the atmosphere plus wind model, 
the resulting fit underestimates the red portion of Si XIV and greatly over-estimates the absorption in the red portion of Fe XXVI. 
In our fitting experiments, we found that adding dynamically broadened re-emission from the same absorbing gas could help achieve better fits to both Si and Fe lines. Although this is true even in the single component model, adding broad emission lines to the wind or the highly redshifted component (Zone 3) only makes these features more pronounced. Adding re-emission, therefore, did not improve the fit of the single component model. Individually, the addition of either a wind absorber (Zone 2) or the highly redshifted absorber (Zone 3) resulted in only marginal improvements to the model. Instead, the model requires re-emission and two additional absorbers in order to adequately describe the data. However, the errors reported in Table \ref{tab:fit} indicate that most parameters for this absorber are poorly constrained. In particular, the negative $1\sigma$ error on the absorbing column (${N}_{He} = {10}_{-9}^{40} \times {10}^{22}$ ${\text{cm}}^{-2}$) approaches unity with the best-fit value and would suggest that the wind is only significant to $1\sigma$. This is in part driven by the complexity of the model that requires a larger change in ${\chi}^2$. Given the tension between the significance suggested by the F-test and the error in ${N}_{He}$, we advise caution when interpreting this result.

The large uncertainty in the outflow velocity of the absorber, however, is due to a local ${\chi}^2$ minimum in a different region of parameter space. The best-fit model reported in Table \ref{tab:fit} fits the bulk of the absorption lines with the absorber corresponding the redshifted disk atmosphere and the blueshifted absorber fits the blue-wing of the Fe XXVI line, only (see Figure \ref{fig:wind}). As can be seen in the middle panel of Figure \ref{fig:redshift}, however, the Si XIV line profile also demonstrates some evidence of complexity in the form of a blue-wing. We performed an alternative fit to Spectrum A in order to capture the blue-wings of both Fe XXVI and Si XIV line profiles, which we report in Table \ref{tab:newfit}. Although the parameters are quite different, statistically, the fits nearly identical. Strikingly, this new fit suggests that the redshift in the atmosphere is much stronger ($\sim 600$ $\text{km}$ $\text{s}^{-1}$), while the blueshift in the ``wind'' component is much lower ($\sim -200$ $\text{km}$ $\text{s}^{-1}$) and is essentially consistent with zero. 
A plausible interpretation is that, in this spectrum, we are probing different regions of the disk atmosphere, where the inner atmosphere is noticeably redshifted while the outer regions are only slightly redshifted.
It is important to note that the redshift of the atmosphere given by the former model is consistent with the fits of Spectrum B, Spectrum $\Gamma$, and the combined A $+$ B $+$ $\Gamma$ Spectrum, while the alternative model gives a much larger redshift ($\sim 600$ $\text{km}$ $\text{s}^{-1}$) and less self-consistency.

The possible presence of a disk wind in UCXBs such as 4U 1916$-$053 can provide key insights about the physical mechanisms that drive disk winds in accreting black holes and neutron stars in general. Magnetic winds in LMXBs are of particular interest as they provide an observational evidence of different magnetic processes mediating mass and angular momentum transfer within the disk. Furthermore, evidence of self-similar winds driven via MHD pressure, possibly as a result of the magnetorotational instability, has been uncovered in GRO J1655-40 \citep{Fukumura2017} and 4U 1630$-$472 \citep{Trueba2019} through their density and velocity structure. 

\begin{table}[t]
\renewcommand{\arraystretch}{1.15}
\caption{Alternative Fit to Spectrum A}
\vspace{-1.0\baselineskip}
\begin{footnotesize}
\begin{center}
\begin{tabular*}{0.47\textwidth}{l   c  c @{\extracolsep{\fill}}  c  }
\tableline
\tableline
Parameter & Zone 1 & Zone 2 & Zone 3 \\
\\ [-3.0ex]
\tableline 
\\ [-3.0ex]

${N}_{He}$ (${10}^{22} {\text{cm}}^{-2}$)&${45}^{+5\ddagger}_{-20}$ & ${20}_{-18}^{+30\ddagger}$ & ${1.3}_{-1.1}^{+15.0}$  \\
log $\xi$&${4.5}_{-0.1}^{+0.3\ddagger}$ & ${4.4}_{-0.5}^{+0.4\ddagger}$ & ${3.7}_{-0.9}^{1.0}$ \\
${v}_{abs}$ ($\text{km}$ $\text{s}^{-1}$) & ${580}_{-550}^{+740}$ & ${-220}_{-1800\ddagger}^{+220\ddagger}$ & ${4200}_{-2600}^{+2800}$  \\
${v}_{turb}$ ($\text{km}$ $\text{s}^{-1}$) & ${70}_{-20\ddagger}^{+430\ddagger}$ & ${300}^{\dagger}$ & ${1000}^{\dagger}$   \\
${\sigma}_{emis}$ ($\text{km}$ $\text{s}^{-1}$)  &${8000}_{-7000\ddagger}^{+12000\ddagger}$ & $-$ & $-$ \\
\\ [-3.0ex]
\tableline 
\\ [-3.0ex]
$\chi^{2}/\nu$ && & 533/488 = 1.09\\

\\ [-3.0ex]
\tableline 
\tableline

\end{tabular*}
\vspace*{-1.0\baselineskip}~\\ \end{center} 
\tablecomments{Parameter values for the alternative fit to Spectrum A. Unlike the fit listed in Table \ref{tab:fit}, Zone 2 captures the blue-wing in both Fe XXVI \emph{and} Si XIV lines. The fact that the data can be described by this alternative fit drives much of the degeneracy when fitting this spectrum.}
\end{footnotesize}
\label{tab:newfit}
\end{table}

Determining whether a wind is driven via magnetic processes has primarily relied on ruling out alternative driving mechanisms via estimates of their launching radii. Although radiation pressure driving (via line driving) can be ruled-out given the high degree of ionization within the gas, thermally driven winds could still arise via Compton heating of the disk surface by energetic photons from the central engine. In principle, this can only occur at radii larger the Compton Radius (or, ${R}_{C}$), where the temperature of the disk surface allows a sizable portion of the gas to exceed the local escape velocity \citep{Begelman1983}, although this limit may extend down to 0.1${R}_{C}$ \citep{Woods1996}. In sources such as GRS 1915$+$105 \citep{Miller2015,Miller2016}, GRO J1655$-$40 \citep{Miller2006a,Miller2008,Miller2015,NeilsenHoman,Kallman2009}, and (among others) 4U 1630$-$472 \citep{Miller2015,Trueba2019}, wind launching radii derived by modeling the photoionzed wind absorption (and, in some cases, wind re-emission) have been well below ${R}_{C}$, suggesting magnetic driving. In the helium rich disk of 4U 1916$-$053, ${R}_{C} = 2 \cdot {10}^{10} \times ({M}_{BH}/{M}_{\odot})/{T}_{C8}$ cm, where the mean molecular weight of \emph{fully ionized} helium gas (${\mu}_{He} \sim 1.3$) makes this limit larger ($\sim2\times$) compared to solar abundances. 
 
There are, however, uncertainties that can significantly limit our ability to measure wind launching radii. In most cases, the lack of an independently measured gas density means that we can only place an upper limit the radius given by photoioinization parameter, $\xi = L/n{r}^2$. This is worsened by our imperfect understanding of the luminosity and shape of the ionizing continuum, as well as poorly constrained distances to these sources. In the case of 4U 1916$-$053 and UCXBs in general, the outermost radius of their small disks (truncated by the orbit of the companion) is itself a useful upper limit on the launching radius of a possible disk wind. Given the orbital period of the X-ray dips and range of possible neutron star masses, we can obtain estimates of this outermost radius through basic Keplerian mechanics, which we can then compare to the Compton radius. Unlike the photoionization radius, this upper limit is model independent and does not depend on the distance, luminosity, or SED shape.

Figure \ref{fig:Rorb} shows a contour plot of the ratio of outermost disk radius over ${R}_{C}$, as a function of neutron star mass and spectrum temperature. The outermost radius is given by ${R}_{outer} \sim {R}_{donor}\times 0.5/{(1+q)}^{1/3}$ \citep{p1967}, 
${R}_{outer} \sim {R}_{donor}\times0.6/(1+q)$, 
where ${R}_{donor}$ is the orbital radius of donor star and q is the binary mass ratio (we assume q $<<1$ in order to establish an upper limit). Our understanding of the broadband continuum is poor within the narrow Chandra energy range, making it difficult to constrain the temperature of disk (at low energies), the temperature of the neutron star, and which of these determines the Compton temperature. Instead of a assuming a single temperature, we chose to plot this ratio for a range of possible temperatures. Our results suggest that the outermost radius of the disk is smaller than the Compton radius, with the largest ratio indicating a disk smaller than $\sim0.2{R}_{C}$. Although $0.1{R}_{C}$ is the commonly used analytical lower limit for thermal driving, only winds launched from the outermost edge of the disk from the low-mass neutron stars and the highest Compton temperatures seem compatible with thermal driving. 
The strict upper limit set by the size of the accretion disk suggests that the possible disk wind found in Spectrum A of 4U 1916$-$053 would likely be magnetic.

\section{Discussion and Conclusions}\label{sec:discussions}

We have analyzed three Chandra/HETG spectra of the short-period neutron-star X-ray binary 4U 1916$-$053.  Two of these are groups of spectra selected from 250~ks of new exposure, and the third was an old archival exposure.  In each case, the bulk of the absorption observed in the steady (non-dip, non-burst) phases is redshifted ($z = 8 \times 10^{-4}$).  It likely arises in an ionized, static, inner disk atmosphere ($R \simeq 1200~GM/c^{2}$), where gravitational red-shifts are small but nonzero.  This interpretation is fully consistent with detailed photoionization modeling including re-emission shaped by local Keplerian motion.  Two of the spectra show evidence of a more strongly red-shifted absorption component, corresponding to $z = 1.4\times 10^{-2}$ and $R \simeq 70~GM/c^{2}$.   One of the spectra shows a clear extended blue wing on the H-like Fe XXVI absorption line, signaling a transient disk wind.  

This disk wind, if real, likely requires some form of magnetic driving.  The orbital period of the X-ray dips indicates that the disk itself is smaller than the Compton radius, suggesting that any disk wind found in this system is likely magnetic \citep{Woods1996,ProgaKallman,Proga2003}. Unlike the disk winds found in most neutron star and black hole systems, then, thermal driving can be ruled out without the uncertainties involved with modeling photoionized absorption. However, alternative fits suggest that the blue wings in both Fe XXVI and Si XIV could, instead, arise from outer portions of the same inner disk atmosphere. 
In this case the data can be fit with two absorbers, the first redshifted by $\sim 600$ $\text{km}$ $\text{s}^{-1}$ and the second compatible with no shift.

In this new Chandra/HETG dataset, a near fivefold increase in total exposure of 4U 1916$-$053, we have uncovered evidence of what is perhaps the largest gravitational redshift ever detected in absorption at $\sim 250$ $\text{km}$ $\text{s}^{-1}$. By comparison, the surface redshifts detected in white dwarfs are of the order of $\sim 35$ $\text{km}$ $\text{s}^{-1}$, with the strongest redshift at 75 $\text{km}$ $\text{s}^{-1}$ \citep{Falcon2010}. 
Archival Chandra/HETG spectra of sources such as XTE J1710$-$281 and AX 1745.6$-$2901 reveal comparable redshifts ($\sim 300$ $\text{km}$ $\text{s}^{-1}$) that have not been previously reported. As with 4U 1916$-$053, these redshifts are significantly higher than the expected relative radial velocity within galaxy and seem to support the interpretation that these redshifts are gravitational. Comparable redshifts in the Chandra/HETG spectrum of MAXI J1305$-$704 suggest that gravitationally redshifted disk atmospheres may also be present in black hole systems, although \citet{Miller2014} make note of issues with this observation that put these results in question. Redshifted absorption features have also been claimed in the spectrum of some AGN. These features, however, are transient, detected at CCD energy resolution (and calibration), and could perhaps be attributed to noise given the quality of the data. Unlike AGN, the redshifted absorption we find in the disk atmospheres of UCXBs is highly significant (4.5 to 5$\sigma$, depending on the test) and, except for hard states in which the gas is over-ionized, persistent. 

These results suggest that UCXBs (as well as sources with slightly longer orbital periods) harbor disk atmospheres with similar properties, where the bulk of the absorption occurs at $R \simeq 1000~GM/c^{2}$, resulting in a detectable gravitational redshift. If this is indeed the case, these atmospheres would form a new class of absorbers in which distance from the compact object could be directly measured with unprecedented accuracy. This also places a hard upper limit on the size of the central engine and would likely require it to be significantly smaller than the radius at which the atmospheric absorption occurs. 

We advise caution when considering these results, especially given the possible alternative explanations for the observed redshifts. In this work, however, we have demonstrated the great wealth of information that has been gained from a significant increase in Chandra exposures of 4U 1916$-$053 and untapped value of studying these sources with Chandra/HETG. We find that additional Chandra exposures of 4U 1916$-$053 could increase the significance of the static lines we find the dips (See Figure \ref{fig:Dips}) and more robustly rule-out radial velocities as the source of redshift without the need of optical or IR observations. The additional exposures could also increase the significance of a wind, if one is present. Further Chandra/HETG observations of sources such as AX 1745.6$-$2901 could also increase the significance of the redshift to the level that we observe in 4U 1916$-$053 and XTE J170$-$281. 

In anticipation of near-future missions such as XRISM, this work further establishes ultra-compact sources such as 4U 1916$-$053 (as well as AX 1745.6$-$2901 and XTE J1710$-$281) as high-priority targets. The dramatic increase in both spectral resolution (near Fe~K) and effective area are ideally suited to perform the analysis described in this work and provide results at a much higher significance with shorter exposures. Our results provide XRISM with clear science goals for these sources, with simple observational signatures (primarily constraining the velocity shift in the atmosphere and during dips) that maximize XRISM's capabilities. In addition, XRISM will be able to confirm whether transient features such as the disk wind and highly redshifted absorber are real and perhaps probe the physics that underlie the disk atmosphere throughout disk.

\clearpage

%----------------------------------------------------------------------

%----------------------------------------------------------------------

%----------------------------------------------------------------------

\end{document}